\documentclass{nsr}

\usepackage{amsmath,graphicx,array}
\usepackage{dcolumn,soul}%

\usepackage{amsthm}
\usepackage[figuresright]{rotating}%
\usepackage{algorithm, algorithmicx, algpseudocode}
\usepackage{listings}%
\usepackage{hyperref}

\makeatletter
\def\uns{\ifmmode\,\else$\,$\fi}%

\makeatother
 
\jvol{XX}
\jnum{X}
\jyear{Year}
\doi{10.1093/nsr/XXXX}
\received{XX XX Year}
\revised{XX XX Year}
\accepted{XX XX Year}

\begin{document}

\dhead{PERSPECTIVE}

\subhead{PHYSICS}

\title{Is ruthenium dioxide altermagnet?}

\author{Alexander A. Tsirlin$^{1,*}$}

\author{Ece Uykur$^{2,*}$}

\author{Oleg Janson$^{3,*}$}

\affil{$^1$Felix Bloch Institute for Solid-State Physics, University of Leipzig, 04103 Leipzig, Germany}

\affil{$^2$Helmholtz-Zentrum Dresden-Rossendorf, Inst Ion Beam Phys \& Mat Res, D-01328 Dresden, Germany}

\affil{$^3$Institute for Theoretical Solid State Physics, Leibniz IFW Dresden, 01069 Dresden, Germany}

\authornote{\textbf{Corresponding authors.} Email: altsirlin@gmail.com, e.uykur@hzdr.de, o.janson@ifw-dresden.de}

\abstract[]{}



\maketitle

Ruthenium dioxide was named as one of the first and most promising altermagnetic candidates with $d$-wave symmetry~\cite{smejkal2022}. Its anticipated band splitting of 1.4 eV along with the presumed magnetic ordering well above room temperature would render this material especially suitable for applications, but later research raised strong doubts, not only on the exact nature of the altermagnetic state, but also about the very existence of magnetic order in this material.

Initial evidence for the altermagnetic nature of RuO$_2$ was mainly based on density-functional theory (DFT) calculations. Experimentally, direct signatures of the $\mathbf k=0$ (altermagnetic) order were inferred from the symmetry-forbidden $(100)$ Bragg peak observed by neutron diffraction~\cite{berlijn2017} and resonant x-ray scattering~\cite{zhu2019}, although polarized neutron experiments showed that this Bragg peak is mainly nuclear in nature~\cite{berlijn2017}. This observation led to a profound discrepancy between the local magnetic moment of about 1\,$\mu_B$, as predicted by DFT, and the \textit{upper} estimate of 0.05\,$\mu_B$ from the neutron experiments.

Addressing reliability of the DFT predictions for RuO$_2$ requires a scrutiny of the underlying computational methodology. The calculations are typically performed with the on-site Coulomb repulsion (Hubbard term) of at least $U=2$\,eV, which mimics correlation effects and stabilizes magnetic order. By removing the Hubbard $U$ and resorting to a standard, uncorrelated DFT calculation, one arrives at a robust nonmagnetic solution. The magnitude of electronic correlations is, therefore, one crucial parameter that controls whether or not RuO$_2$ is altermagnet.

On the experimental side, photoemission experiments~\cite{liu2024} as well as optical spectroscopy~\cite{wenzel2025} as direct probes of the electronic structure lend firm support to the nonmagnetic scenario (Fig.~1b,c). The uncorrelated, nonmagnetic DFT calculation allows a very good agreement for both experimental band dispersions and optical conductivity. Additionally, the intraband part of the optical response (Drude spectral weight and plasma frequency) serves as a useful gauge of the correlation effects and clearly indicates their weakness in RuO$_2$~\cite{wenzel2025}. 

Bulk RuO$_2$ would be altermagnetic only in the presence of sufficiently strong electronic correlations, which however are not seen in this material experimentally. Indeed, recent studies show that the $(100)$ Bragg peak reported in Ref.~\cite{berlijn2017} as an evidence of magnetism was very likely an artifact of double diffraction in the neutron experiment~\cite{kessler2024} (Fig.~1d), whereas the appearance of this peak in resonant x-ray scattering is due to an anisotropic charge distribution in the material~\cite{occhialini2025}. Various experimental probes reported on bulk RuO$_2$ over the last two years all demonstrate striking consistency with the nonmagnetic solution from the uncorrelated DFT calculation and fail to see features that would be expected in altermagnetic RuO$_2$. Of particular note is muon spectroscopy that boasts remarkable sensitivity to any magnetic order by detecting weak magnetic fields on the order of 0.1\,G, but even this method failed to resolve any signatures of magnetism in the bulk of ruthenium dioxide~\cite{hiraishi2024}.

\begin{figure*}[!t]
\includegraphics[width=14cm]{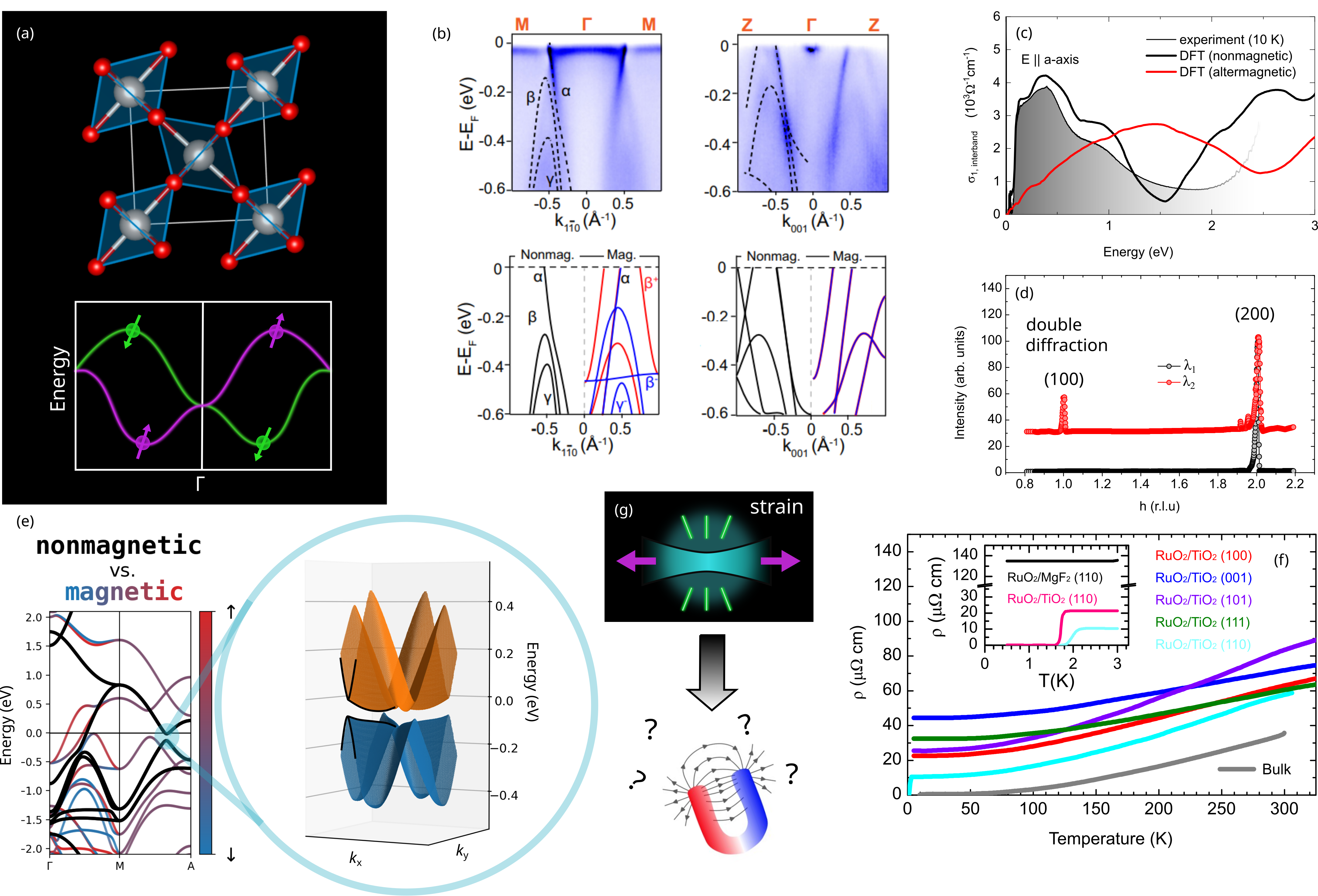}
\caption{
(a) Non-symmorphic crystal structure of RuO$_2$ and altermagnetic band splitting. (b,c) Photoemission~\cite{liu2024} and optical~\cite{wenzel2025} spectroscopies confirm the absence of band splitting in bulk RuO$_2$. (d) Changing neutron wavelength in the diffraction experiment leads to a suppression of the $(100)$ Bragg peak and points to double diffraction as its origin~\cite{kessler2024}. (e) Nonmagnetic band structure of RuO$_2$ features a Dirac nodal line near the Fermi level. (f) High residual resistivity of thin films~\cite{uchida2020,ruf2021,wang2025b} compared to bulk~\cite{wenzel2025} indicates an increased concentration of defects that may also give rise to magnetism observed in some of the thin films. (g) Application of strain is one potential route for making RuO$_2$ magnetic. Panel (b) is reprinted with permission from Ref.~\cite{liu2024}, copyright 2024 by the American Physical Society.
}
\end{figure*}

Is RuO$_2$ a mundane paramagnetic metal? By all means not. The thorough characterization of this material uncovered interesting features of its electronic structure, most notably, the presence of an extended Dirac nodal line slightly below the Fermi level~\cite{wenzel2025,jovic2018} (Fig.~1e). Such a nodal line has direct implications for the Hall response and spin transport. Systematic studies of the latter, with varying directions of spin and charge currents, conclude that promising transport properties of RuO$_2$ and especially the spin torque reported therein can be ascribed to the anisotropic spin Hall effect~\cite{wang2025} and should be thus relativistic in nature, in contrast to the nonrelativistic scenario of altermagnetic band splitting.

An interesting question at this juncture is whether RuO$_2$ can be made magnetic and altermagnetic by applying pressure or strain, introducing defects, or varying oxygen stoichiometry~\cite{smolyanyuk2024}. One good news here is that many of these modifications are already known from the previous literature. Thin films of RuO$_2$ have been grown for various orientations on many different substrates, often for applications in catalysis, whereas defect physics is central to the low-temperature upturn in the resistivity, used in thermometry applications of RuO$_2$. The downside is that any of these tools is unlikely to render RuO$_2$ strongly correlated and restore the altermagnetic scenario with the large band spin splitting anticipated in the initial publications on the basis of DFT. 

Emergence of superconductivity with $T_c$ up to 2\,K in RuO$_2$ thin films grown on TiO$_2$ (110) and MgF$_2$ (110) offers the most crisp example of strain modification (bulk RuO$_2$ shows no traces of superconductivity)~\cite{uchida2020}. Thin films grown on other substrates are non-superconducting, whereas their possible magnetism remains controversial. Spin-transport experiments and x-ray magnetic linear dichroism signals~\cite{zhang2025} are suggestive of the magnetically ordered state. On the other hand, attempts of detecting an excitation mode of this putative magnetic order were so far unsuccessful. The magnetic response of RuO$_2$ films grown on ferromagnetic substrate, the setting often used in spin-transport experiments, was interpreted as a proximity effect, with no intrinsic magnetism of RuO$_2$ itself~\cite{abel2025}. Moreover, low-energy muon spectroscopy, which is geared toward probing magnetism of thin films, reported no signatures of magnetic order even in thin-film RuO$_2$~\cite{kessler2024}. One should also keep in mind that thin films typically feature a much higher residual resistivity compared to the bulk (Fig.~1f), and defects inherently present in the films may be one cause for the weak signatures of magnetism. The abundance of different substrates and orientations available for RuO$_2$ films necessitates further systematic studies, ideally with different methods applied to the same sample and with the focus on direct probes of internal magnetic fields using muon, M\"ossbauer, and resonance spectroscopies. 

In summary, recent research gives overwhelming evidence for the absence of magnetism in bulk RuO$_2$. This material is not as easily accessible room-temperature altermagnet as it was hoped for. On the other hand, available experimental avenues of material modification, along with the theoretical understanding of its proximity to electronic instabilities, leave the question posed in the title of this Perspective essentially open. RuO$_2$ has the right symmetry for altermagnetism, but the suitable routes for making it altermagnetic remain to be found. An important caveat on the theoretical side is that DFT calculations tend to produce ambiguous results for this material. The exact computational methodology should be carefully benchmarked against the extensive experimental information that has been accumulated for bulk RuO$_2$, before useful predictions for strained or otherwise modified RuO$_2$ can be made. An equally important caveat on the experimental side is that every single method can be prone to errors in identifying a material as magnetic or nonmagnetic. Only a combination of scattering techniques and local probes of internal magnetic field would allow a conclusive classification of a suitably modified RuO$_2$ as magnetically ordered and altermagnetic. Finally, spin Hall effect in RuO$_2$ does not rely on altermagnetism. Many of the potential applications of this material do not require it to be altermagnet.


\end{document}